# Spatial mixed models for assessing environmental exposure effects on the microbiome

Sooran Kim[1], Chan Wang[1], Soyoung Kwak[2,3], Fares Darawshy[4,5,6], Alexander Bain[4], Leopoldo N. Segal[4], Jiyoung Ahn[2,3], Huilin Li[1*]

[1]Division of Biostatistics, Department of Population Health, New York University School of Medicine, New York, NY, USA

[2]Department of Population Health, New York University Grossman School of Medicine, New York, NY, USA

[3]NYU Laura and Isaac Perlmutter Cancer Center, New York, NY, USA

[4]Department of Medicine, Division of Pulmonary, Critical Care, and Sleep Medicine, NYU Langone Medical Center, New York, NY, USA

[5] The Institute of Pulmonary Medicine, Hadassah Medical Center, Jerusalem, Israel

[6] Faculty of Medicine, Hebrew University of Jerusalem, Jerusalem, Israel

[*]Correspondence: Huilin.Li@nyulangone.org

Emails: Sooran Kim: Sooran.Kim@nyulangone.org, Chan Wang: Chan.Wang@nyulangone.org, Soyoung Kwak: Soyoung.Kwak@nyulangone.org, Fares Darawshy: Fares.Darawshy@nyulangone.org, Alexander Bain: Alexander.Bain@nyulangone.org, Leopoldo Segal, Leopoldo.Segal@nyulangone.org, Jiyoung Ahn: Jiyoung.Ahn@nyulangone.org, Huilin Li: Huilin.Li@nyulangone.org

## Abstract

**Background:** The influence of environmental exposures, such as air pollution, on human health has become increasingly recognized. A growing body of evidence suggests that the microbiome may mediate these effects, explaining the relationship between the environment and host biology. However, the impact of environmental exposures on the microbiome is not yet fully understood, and statistical modeling in this context is challenged by complex dependency structures. In particular, microbiome data exhibit spatial dependencies across sampling regions as well as ecological correlations among microbial taxa, which, if ignored, can substantially reduce detection power, leading to missed true signals.

**Results:** We introduce a novel spatial mixed modeling framework for microbiome data that accounts for both region-level spatial dependency and taxon-level ecological dependency using conditional autoregressive priors. Through simulations, we demonstrate that this framework outperforms existing methods that ignore such dependencies, by achieving high detection power in feature selection while maintaining low false positive rates and reduced mean squared error in estimation. Applied to two real studies-data from Food and Microbiome Longitudinal Investigation study and lung microbiome dataset-with fine particulate matter ($PM_{2.5}$) exposures, our model identified genera*,* which are known to be involved in pollution-related health outcomes, as well as novel taxa that may mediate host responses to air pollution.

**Conclusions:** This novel approach offers a powerful and flexible tool for uncovering biologically meaningful associations in complex environmental data.

## Background

Environmental exposures, such as air pollution and chemical contaminants, have long been recognized as critical determinants of human health. Many of these exposures vary spatially, creating geographic patterns that can influence population-level health outcomes. At the same time, high-dimensional molecular data–such as microbiome profiles, metabolomics, transcriptomics, and epigenomics–are available through high-throughput technologies from the large-scale population studies. These data provide unprecedented opportunities to investigate how environmental factors affect biological systems. However, traditional statistical methods for epidemiologic and molecular association analysis often struggle to account simultaneously for high-dimensional molecular features, structured spatial dependencies, and feature-level biological correlations. This gap highlights the need for novel statistical and computational approaches, which can integrate spatially structured environmental information with high-throughput molecular data.

As a motivating example, the association between human microbiome and environmental factors, such as fine particulate matter ($PM_{2.5}$), has received increasing attention for its relevance to various health issues [1, 2]. Previous studies have shown that $PM_{2.5}$ exposure can affect microbiome composition, potentially altering gut microbial diversity [3]. Such exposure-related microbial changes may contribute to health problems such as diabetes risk [2] or mental health [4]. In addition to the gut microbiome, the lung and airway microbiome represent another biologically relevant setting for studying $PM_{2.5}$-associated microbial variation. Several studies of the airway or upper-respiratory microbiome have reported associations between particular matter exposure, including $PM_{2.5}$, and respiratory microbial composition, diversity, or microbiome-mediated respiratory health outcomes [5-8]. A recent review further summarized evidence that air pollution can disrupt the respiratory microbiome [9]. Therefore, it is crucial to identify specific microbial species influenced by $PM_{2.5}$ for understanding its impact on human health.

In such a microbiome study, exploring the dependence, arising from either geographic factors or between the microbiome taxa, plays a crucial role. However, such dependencies have been largely overlooked in previous work. First, microbiome is often impacted due to the geographically varying covariates. While some of this variation could be explained by environmental exposures such as $PM_{2.5}$, residual spatial structures may remain unexplained. Second, microbiome taxa are not independent; they often exhibit positive or negative correlations due to ecological interactions, shared functions, or compositional constraints [10, 11], and these correlations need to be accounted for in the model.

To address such dependencies, a common approach is the use of mixed models. While mixed models have been widely applied in microbiome research, most studies however focus on temporal correlations in longitudinal data and typically overlook the spatial dependence [12-15]. On the other hand, Martin, Witten [16] considered within-sample correlation, they did not model between-taxon dependencies, while Lee, Coull [17] used an unstructured covariance for random effects to capture taxa correlations, which gives maximal flexibility. However, such complexity of the model might be unnecessary, due to the high cost of computation.

To fill these gaps, we develop a novel modeling framework to identify specific microbial species influenced by environmental exposures, while explicitly incorporating spatial dependencies across sampling regions and ecological correlations among microbial taxa. To model the sequencing read counts, we adopt a zero-inflated Poisson (ZIP) distribution, which is widely used in microbiome research [17-19]. We also include the core principle of ANCOM-BC, which utilizes a log-linear model to estimate the unknown sampling fraction [20]. In microbiome studies, the observed abundances are constrained by the library size, leading to technical biases where the sequencing depth varies across samples. ANCOM-BC addresses this by introducing a sample specific bias term–the sampling fraction–into the model. By treating this fraction as a random effect within our framework, we can effectively capture between-sample variation in sequencing depth. This approach allows us to separate true biological variation in microbial abundances from technical variation introduced by unequal sequencing effort across samples. On the other hand, unlike existing approaches, our model includes a conditional autoregressive (CAR) prior to account for region-level residual spatial dependence not explained by environmental exposure. Additionally, we incorporate another CAR prior to capture correlation among taxa. In this way, our model explicitly incorporates dependencies at both the region- and taxon-levels.

The estimation and inference in our model pose substantial challenges due to several factors. First, microbiome data often exhibit zero-inflation, where a large proportion of counts are zeros. This makes standard count models such as the Poisson inadequate. Second, the data are typically high-dimensional, with hundreds of taxa across many subjects, which further complicates model fitting. Third, spatial priors such as CAR introduce spatial structure, which adds complexity because the covariance matrix is no longer diagonal and cannot be simplified as in models assuming independence.

To handle these, we employ the Integrated nested Laplace approximation (INLA), which provides an efficient Bayesian framework for latent Gaussian Markov random field models [21]. INLA approximates marginal posterior distributions, while avoiding the computational burden of full Markov Chain Monte Carlo (MCMC) sampling. In addition, it accommodates various types of random effects, including CAR priors, which are central to our model. Thus, our methodology can be efficiently implemented using the R-INLA package [21].

Although this article focuses on environmental exposure-microbiome associations, the proposed framework is not limited to microbiome data. It can, in principle, be applied to other high-dimensional count data measured on spatially indexed subjects. In such settings, microbial taxa may be replaced by other count-based biological features such as RNA-seq data. Across these applications, spatial autocorrelation among subjects and dependence among high-dimensional features can pose substantial challenges for standard association analyses. Our framework provides a unified approach for addressing these issues in high-dimensional count data.

To illustrate these challenges in a concrete setting and motivate our proposed spatial mixed modeling framework, we first examine one of our motivating applications, the Food and Microbiome Longitudinal Investigation (FAMiLI) study.

**Motivating Example: Food and Microbiome Longitudinal Investigation (FAMiLI)**

In the following application, we analyze 16S rRNA microbiome sequencing data from a subset of 825 participants enrolled in the FAMiLI study, a U.S.-based cohort (n=13,101) representing diverse racial and nativity backgrounds [22]. Air pollution exposure data were obtained from the dataset developed by Rahman and Thurston [23].

With a focus on New York City (NYC), which provides a well-connected neighborhood structure, our primary goal is to investigate the associations between gut microbial taxa and the environmental exposure $PM_{2.5}$. Prior to the main analysis presented in the Results section, we examine the spatial and ecological dependencies present in the datasets.

To evaluate spatial autocorrelation, we conducted Moran's I test, based on data from 381 subjects across 84 postal codes (zip codes in the U.S. context) at NYC. First, we performed Moran's I tests on alpha-diversity measures, which represent community-level microbiome diversity. For these tests, a subject-level neighborhood matrix was defined such that two subjects were considered neighbors if they resided in the same or adjacent regions. All four alpha-diversity measures (observed amplicon sequence variants (ASVs), Shannon index, Simpson's index, Inverse Simpson index) yielded p-values less than 0.05, indicating significant spatial dependence. The spatial distributions of the average alpha-diversity measures across regions are given in Additional file 1: Figure S1.

Second, we investigated spatial structure in microbiome at taxonomical level. A multivariate spatial autocorrelation test (*multispati.randtest* in R) was conducted, considering all taxa simultaneously rather than individually. Prior to the test, we applied a minimum prevalence filter of 10%, resulting in 212 operational taxonomic units (OTUs) included for further analysis. Subsequently, the centered log-ratio (CLR) transform (a small pseudocount (0.01) was added to handle zeros) was then applied to account for variation in sequencing depth across sample. Principal component analysis (PCA) was performed on the CLR-transformed data to reduce dimensionality, and a multivariate spatial autocorrelation test was conducted on principal component scores to assess spatial autocorrelation of the microbiome community across regions. The test revealed a significant spatial structure (p-value = 0.001 based on 999 permutations). Figure 1 presents the spatial distributions of the average values of the first and second principal components scores (PC1 and PC2, respectively), aggregated at the postal code level. This approach was preferred over per-taxon tests which can produce inflated results due to compositional dependencies.

Furthermore, we evaluated the correlation among taxa. Several approaches can be used to assess these correlations, including phylogenetic relationships or functional similarity. In the phylogenetic approach, a distance matrix such as the cophenetic distance matrix from a phylogenetic tree can be computed. This reflects the evolutionary relatedness among taxa. This distance matrix is presented in Additional file 1: Figure S2 (a). As an illustrative alternative, we also considered functional similarity, which captures the similarity of gene content across taxa [24]. Even though taxa belong to different taxonomic groups, they can have similar functional roles, capturing shared biological meaning. For example, populations may exhibit different taxonomic profiles due to diet or cultural differences, yet maintain similar underlying functionality. Furthermore, functionally similar taxa might be likely to respond to environmental factors in similar ways. Therefore, functional similarity can enhance model stability and biological interpretability. Functional similarity at the genus level was defined as $1 - d_{ik}$, where $d_{ik}$ represents taxon-taxon functional distance between taxon $i$ and taxon $k$ as described in Tian, Wang [24]. Specifically, $d_{ik}$ was derived from the genomic content network (GCN) which is a bipartite graph connecting microbial taxa to the genes within their genomes. The GCN was constructed using metagenomic data from all available healthy samples in the curatedMetagenomicData dataset [25]. Suppose that the microbiome data have $p$ taxa and $q$ functional genes. The GCN is represented by a matrix $\boldsymbol{G}$, where a non-negative integer $G_{jg}$ indicates the copy number of gene $g$ in the genome of taxon $j$. The functional distance $d_{ik}$ was defined as the weighted Jaccard distance between genomes of taxon $i$ and taxon $j$: $d_{ik} = 1 - \sum_{\mathrm{g}} \min(G_{ig}, G_{kg}) \,/\, \sum_{\mathrm{g}} \max(G_{ig}, G_{kg})$. Since only a subset of taxa was mapped onto the genome content network, functional similarities for taxa that could not be projected were conservatively set to zero. Additional file 1: Figure S2 (b) displays the heatmap of the functional similarity matrix among 212 OTUs, revealing clear clusters of taxa with shared functional profile. We emphasize that this matrix was used to illustrate one possible construction of taxon-level dependence, rather than as a definitive or fully validated representation of microbial functional relationships. Overall, these results show the need for a novel modeling framework, incorporating such dependencies.

The proposed framework follows a structured workflow that integrates data preprocessing, dependence assessment, model specification, Bayesian estimation, and feature selection. We first described the preprocessing steps and presented diagnostic procedures for assessing spatial dependence across regions and ecological dependence among taxa in this section. Building on these components, we will introduce the proposed model and describe the downstream inferential procedures, including feature selection based on local false discovery control. An overview of the complete workflow is provided in Figure 2.

## Methods

In this section, we describe the proposed ZIP mixed model, prior distributions for model parameters, and implementation.

### Zero-inflated Poisson Mixed Models

We begin this section with introducing the rationale of ANCOM-BC [20, 26]. Their framework accounts for sample-specific bias by introducing sampling fraction terms $\theta_i$, which are treated as nuisance parameters. These terms are estimated as residuals from a log-linear model; $\hat{\theta}_i = \sum_{j=1}^{J} \left( y_{ij} - \beta_j^\top x_i \right) / J$, where their underlying assumption is all taxa in the same sample are assumed to be scaled by the same sampling fraction which introduces compositional bias. Thus, estimating this shared sampling fraction across all taxa allows for proper de-biasing of observed data. In this paper, we adopt this idea and adapt the sampling fraction as a random effect. This formulation is justified by noting that the residual estimator $\hat{\theta}_i$ in ANCOM-BC can be viewed as the posterior mode (i.e., maximum a posterior, MAP) estimator of a Gaussian random effect $\theta_i \sim N(0, \sigma_\theta^2)$ in a log-linear model. Specifically, for sample $i = 1, \cdots, n$, and taxon $j = 1, \cdots, J$, suppose we model the log-counts as

$$y_{ij} = \theta_i + \beta_j^\top x_i + \epsilon_{ij}, \quad \epsilon_{ij} \sim N(0, \sigma_\epsilon^2),$$

with $\theta_i \sim N(0, \sigma_\theta^2)$. The posterior mode of $\theta_i$ is obtained by maximizing

$$l(\theta_i) \propto -\frac{1}{2\sigma_\epsilon^2} \sum_{j=1}^{J} \left(y_{ij} - \theta_i - \beta_j^\top x_i\right)^2 - \frac{\theta_i^2}{2\sigma_\theta^2}.$$

Differentiating and setting to zero,

$$\frac{1}{\sigma_\epsilon^2} \sum_{j=1}^{J} \left(y_{ij} - \theta_i - \beta_j^\top x_i\right) - \frac{\theta_i}{\sigma_\theta^2} = 0,$$

which gives the solution

$$\hat{\theta}_i^{MAP} = \frac{\sum_{j=1}^{J} \left(y_{ij} - \beta_j^\top x_i\right)/J}{1 + \sigma_\epsilon^2/(J\sigma_\theta^2)}.$$

As $\sigma_\theta^2 \to \infty$, the denominator approaches 1, and

$$\hat{\theta}_i^{MAP} = \frac{1}{J} \sum_{j=1}^{J} \left(y_{ij} - \beta_j^\top x_i\right),$$

which coincides exactly with the residual estimator in ANCOM-BC. For finite $\sigma_\theta^2$, the MAP estimate corresponds to a shrinkage version of the ANCOM-BC residual, since the log-likelihood includes a ridge penalty on $\theta_i$. Thus, ANCOM-BC residual estimates can be seen as a special case of posterior modes in a random effect model. To integrate the principles of ANCOM-BC into our proposed model, we incorporate a sample-level random effect $v_{li}$, which represents the sample-specific sampling fraction.

Now, we propose a model to describe the microbiome, which is influenced by environmental exposure as well as dependence between taxa. Let each spatial location $l = 1, \cdots, L$ contain $n_l$ subjects, indexed by $i = 1, \cdots, n_l$. Further, assume that there are $J$ microbial taxa, indexed by $j = 1, \cdots, J$. Then, the sequencing read count of taxon $j$ for subject $i$ at location $l$, denoted as $\{M_{lij}: l = 1, \cdots, L,\ i = 1, \cdots, n_l, j = 1, \cdots, J\}$, is modeled using a ZIP framework:

$$M_{lij} \sim \begin{cases} 0, & \text{with probability } p_{lij} \\ \text{Poisson}(\lambda_{lij}), & \text{with probability } 1 - p_{lij} \end{cases} \quad \textbf{(1)}$$

where the Poisson mean is linked to covariates and random effects through

$$log(\lambda_{lij}) = \beta_{0j} + A_{li}\beta_{A,j} + X_{li}\beta_{X,li} + u_l + v_{li} + w_j.$$

Here, $A_{li}$ represents the exposure to subject $i$ at location $l$, and $X_{li}$ denotes an additional covariate vector that can accommodate both continuous and categorical variables. The coefficient $\beta_{A,j}$ is the primary parameter of interest in this study, representing the effect of exposure on the abundance of taxon $j$ after adjusting for covariates and random effects. Further, $u_l$, $v_{li}$, and $w_j$ are region-, sample (or subject)-, and taxon-level random effects, respectively. The region-level random effect $u_l$ captures spatial variation across regions, and the taxon-level random effect $w_j$ accounts for correlated behavior across taxa. The sample-level random effect $v_{li}$ is assumed to follow a normal distribution with mean zero and variance $\sigma_v^2$. This accounts for between-sample variation such as sampling fraction. Notably, the index $l$ denotes the region associated with subject $i$. Thus, $v_{li}$ represents a sample-specific effect nested within a spatial region. This allows the model to account for variation in sequencing depth at the sample-level, while maintaining a hierarchical structure that is nested within a spatial region.

## Integrated nested Laplace approximation (INLA)

For estimation and inference, we adopt a Bayesian framework, specifically INLA [21]. Classical methods, such as maximum likelihood estimation or expectation-maximization algorithm are not well-suited for this framework. Incorporating zero-inflation, multiple layers of random effects including spatial random effects, and hierarchical structure simultaneously leads to a highly non-standard likelihood function, which is difficult to optimize and may result in unstable estimates with high computational cost.

The Bayesian framework accommodates these complexities through flexible prior specifications. In particular, we use the INLA, a method for latent Gaussian Markov random field models. INLA is a well-established and computationally efficient alternative to MCMC methods. Rather than sampling from the full joint posterior distribution, INLA uses approximations to compute the marginal posterior distributions of parameters. On the other hand, MCMC methods construct a Markov chain whose stationary distribution is the posterior, which requires careful convergence assessment and discarding of burn-in iterations. This aspect of INLA leads to substantial computational gains. Moreover, the R-INLA package provides a convenient and accessible implementation, allowing for straightforward applications of complex spatial and hierarchical models.

## Prior Specification

We specify priors for the unknown parameters to complete the Bayesian formulation in this section. For the intercept parameters $\beta_{0j}$, we use diffuse normal priors, while for the exposure effects $\beta_{A,j}$, we assign $\beta_{A,j} \sim N(0, \tau_A^{-1})$ for $j = 1, \ldots, J$, where the precision $\tau_A$ follows a penalized complexity (PC) prior. The PC prior framework penalizes model complexity by shrinking toward a base model, which in this case corresponds to $\beta_{A,j} = 0$. It is specified through a probability statement of the form $P\left(\tau_A^{-1/2} > U\right) = \alpha$, which reflects prior belief about the size of the standard deviation. This construction favors the base model and induces shrinkage of the exposure effects, thereby acting as a continuous shrinkage prior. As a result, it indirectly facilitates feature selection by shrinking many coefficients toward zero while allowing non-negligible effects to remain [27, 28].

For region-level random effect $u_l$, we employ a Leroux CAR prior to capture regional spatial dependence [29]. We denote it as CAR($\rho_u$, $\sigma_u^2$):

$$\boldsymbol{u} = (u_1, \cdots, u_L)^\top \sim MVN_L\left(\boldsymbol{0}, \sigma_u^2\left((1-\rho_u)\boldsymbol{I_L} + \rho_u \boldsymbol{R_{region}}\right)^{-1}\right),$$

where $\boldsymbol{I_L}$ denotes the identity matrix with dimension $L$, and $\boldsymbol{R_{region}} = [R_{ll'}]_{l,l'}$ defined as

$$R_{ll'} = \begin{cases} m_l, & l = l' \\ -I(l \sim l'), & l \neq l' \end{cases}$$

Here, $m_l$ denotes the number of neighbors of region $l$, $I(l \sim l')$ is an indicator function equal to 1 if region $l$ and $l'$ are neighbors, and 0 otherwise. The parameters $\sigma_u^2(> 0)$ and $\rho_u(\in (0,1))$ quantify overdispersion and spatial dependence, respectively. A neighborhood structure based on geographic adjacency can be used to capture spatial relationship among regions. We chose a CAR prior for the region-level random effects $u_l$ because our goal is to explore the relationship between microbiome and environmental exposure, rather than to optimize predictive performance. Compared to Gaussian process priors, CAR priors are more computationally efficient due to the sparsity of the neighborhood matrix. Among CAR formulations, we selected the Leroux CAR model, which is a generalization of the independent and intrinsic CAR models and yields a proper distribution of $\rho_u$. This parameterization not only ensures model identifiability but also provides greater interpretability, as the spatial dependence parameter $\rho_u$ can be directly understood as the degree of spatial correlation.

A Leroux CAR prior is also placed on the taxon-level random effect $w_j$, denoted as CAR($\rho_w$, $\sigma_w^2$), and defined as

$$\boldsymbol{w} = \left(w_1, \cdots, w_J\right)^\top \sim MVN_J\left(\boldsymbol{0}, \sigma_w^2\left((1-\rho_w)\boldsymbol{I_J} + \rho_w \boldsymbol{R_{micro}}\right)^{-1}\right),$$

where $\boldsymbol{I_J}$ is the identity matrix of dimension $J$, and $\boldsymbol{R_{micro}}$ is defined analogously to the regional-level matrix $\boldsymbol{R_{region}}$ used for $u_l$. Unlike geographic regions, taxa do not possess explicit spatial coordinates, making Gaussian process priors unsuitable. Instead, we construct $\boldsymbol{R}_{micro}$ using biological relationships among taxa, such as phylogenetic information or other biologically motivated similarity measures. In the phylogenetic approach, a phylogenetic tree is used to compute pairwise distances among taxa. For example, we can employ the cophenetic distance, which quantifies the branch-length distance between pairs of taxa at the tips of a phylogenetic tree and therefore provides a natural measure of evolutionary relatedness [30]. We then apply a threshold to convert this distance matrix into a binary neighborhood matrix. More generally, for any biologically motivated similarity measures, a binary neighborhood matrix can be obtained by thresholding the corresponding similarity matrix, such that pairs of taxa with similarity above a prespecified threshold are treated as neighbors. Consequently, $\boldsymbol{R}_{micro}$ is derived from a binary neighborhood matrix. While several thresholding strategies exist, we adopt a criterion in simulation studies that ensures each taxon has at least 5% of the total number of taxa as neighbors. The threshold provides a balance between maintaining sufficient connectivity for information sharing and preserving sparsity to avoid introducing spurious correlations. A 5% cutoff is adopted as a heuristic to achieve this balance, although formal standards may vary depending on the method and context. Additionally, we assign Beta priors for the spatial dependence parameters $\rho_u$ and $\rho_w$. For the taxon-level spatial dependence parameter $\rho_w$, the choice of prior can be guided by biological knowledge. For example, if the moderate correlation among taxa is expected, a Beta$(2,2)$ prior, which centers at 0.5, is appropriate. If relatively strong correlation is anticipated, a Beta$(5,1)$ prior may be used, reflecting a left-skewed distribution favoring higher values. PC priors were assigned to the precision parameters $\tau_u = \sigma_u^{-2}$, $\tau_v = \sigma_v^{-2}$ and $\tau_w = \sigma_w^{-2}$.

## Simulation Studies

**Simulation Design.** We first fit the proposed model to 600 simulated data sets, with 100 datasets generated under each of the six scenarios that combined two dependence structures (moderate and strong) with varying number of taxa $J \in \{100, 200, 300\}$. The number of truly significant taxa was set to 6, 11, and 16 taxa for $J = 100, 200$, and $300$, respectively. In the Results section, Scenarios I-III correspond to moderate dependence with $J = 100, 200$, and $300$, respectively, while Scenarios IV-VI indicate strong dependence with the corresponding values of $J$.

To ensure realism, parameter values were set based on the estimates obtained from the FAMiLI dataset, which includes 359 individuals across 82 locations. Specifically, fitting a simple CAR model (i.e., CAR($\rho_A, \sigma_A^2$)) to $PM_{2.5}$ yielded $\hat{\rho}_A = 0.9$ and $\hat{\sigma}_A^2 = 1/30$. Using a neighborhood matrix defined by functional similarity matrix, fitting our model **(1)** to the data provided estimates for other variance components; $\hat{\sigma}_v^2 = 0.5$, $\hat{\rho}_u = 0.7, \hat{\tau}_u = 26$, $\hat{\rho}_w = 0.6$, $\hat{\tau}_w = 2$.

Based on these values, we considered a $9 \times 9$ regular lattice wrapped on a torus. The number of subjects per location was mostly one or two, with a few locations containing more than five subjects (Additional file 1: Table S1). The total number of subjects was set to 377. To reflect the behavior of $PM_{2.5}$, the spatially dependent covariate $A_{li}$ was generated from a Leroux CAR model with $\rho_A = 0.9$ and $\sigma_A^2 = 1/30$, and no additional covariates were included for conciseness (i.e., no $X_{li}$). The values of $\beta_{A,j}$ were also set to estimates from FAMiLI (Additional file 1: Table S2). For the sampling fraction, we fixed $\sigma_v^2 = 0.5$. Under moderate dependence, we used the real data-based values; $\rho_u =$

$0.7, \tau_u = 26, \rho_w = 0.6$, $\tau_w = 2$. For strong dependence, we specified a higher-dependence setting $\rho_u = 0.8, \tau_u = 26$, $\rho_w = 0.8$, $\tau_w = 10$.

Baseline abundance tables were generated using Poisson lognormal models (*sim_plnm* in R) to reproduce the distributions observed in FAMiLI. In this approach, zeros arise naturally from the Poisson sampling of low latent means, without requiring an explicit zero-inflation parameter. These were log-transformed, modified to incorporate covariate and random effect contributions, exponentiated, and then rounded to yield integer count data.

The region-level adjacency structure was defined using a rook-contiguity scheme on a $9 \times 9$ lattice, with toroidal boundary conditions to avoid edge effects. The taxon-level neighborhood structure was derived from a simulated phylogenetic tree generated using *rtree* function (*ape* package in R). Cophenetic distances were computed from the tree and used to construct a binary adjacency matrix by connecting each taxon to its nearest neighbors in phylogenetic space, such that each taxon has at least 5% of the total number of taxa as neighbors, as discussed in the Prior Specification section.

Additionally, we evaluated the robustness of the proposed method under model misspecification. In this extra scenario, microbiome count data were generated from a negative binomial distribution under a representative setting with moderate dependence and $J = 200$. Baseline taxon relative abundances and sample library sizes were derived from the FAMiLI dataset, and taxon-specific negative-binomial dispersion parameters were obtained using the mean-variance relationship: $var(Y) = \mu + \mu^2/size$, where $\mu$ represents the mean abundance. Known exposure effect was then added on the log-mean scale, together with random effects. Counts were generated from the resulting negative-binomial mean and dispersion parameters. This scenario evaluates whether the proposed model remains effective when the observed counts arise from a non-ZIP distribution while preserving known ground-truth exposure effects for comparison.

**Specification of Hyperparameters.** Following the Prior Specification section, we specify prior distributions for the hyperparameters. For the intercept parameters $\beta_{0j}$, we specified the variance of the normal prior to 1,000 for $j = 1, \ldots, J$. For the precision parameter $\tau_A$, PC prior defined with $U_A = 0.5$ and $\alpha_A = 0.01$ was used to reflect the belief that most exposure effects are small in magnitude, while allowing occasional moderate effects. For the precision parameters $\tau_u$, $\tau_v$, and $\tau_w$, PC priors were assigned with $U_{uvw} = 2$ and $\alpha_{uvw} = 0.05$. This choice reflects a prior belief that large standard deviations are unlikely while still allowing flexibility when supported by the data. For the dependence parameter $\rho_u$, we assigned a Beta$(1,1)$ prior. For $\rho_w$, we used a Beta$(2,2)$ prior under the moderate dependence scenarios, and a Beta$(5,1)$ prior under the strong dependence scenarios.

We conducted sensitivity analyses under a representative moderate scenario with $J = 200$ to assess the sensitivity of inference for the exposure effects $\beta_{A,j}$. Specifically, we varied the PC prior governing the shrinkage of $\beta_{A,j}$ by considering $U_A \in \{0.25, 0.5, 1\}$ with $\alpha_A = 0.01$, corresponding to stronger to weaker shrinkage. These settings are referred to as $\beta$ low, baseline, $\beta$ high, respectively. In addition, we examined an alternative specification for the precision parameters $\tau_u$, $\tau_v$, and $\tau_w$, by setting $U_{uvw} = 1$, instead of $U_{uvw} = 2$, while keeping $\alpha_{uvw} = 0.05$; this setting is referred to as variance shrink.

**Assessment Criteria.** For comparison, we implemented an independent model following ANCOM-BC [20] in each dataset. This independent model allows us to assess the contribution of incorporating both region- and taxon-level random effects. In particular, we focus on ANCOM-BC rather than ANCOM [31], as ANCOM-BC provides improved bias correction and statistical properties while addressing compositional effects more rigorously. We additionally considered MaAsLin with a random effect for location, representing a mixed effects regression model that accounts for sample-level clustering but does not model spatial random effects or taxon-level dependence [15]. Other existing methods were not included, as they are designed for different study designs (e.g., longitudinal data) or different distributional assumptions (e.g., negative binomial), and do not explicitly model region- and taxon-level dependencies.

For each simulation, we evaluated feature selection performance for the considered methods. For the proposed model, we adopted the local false discovery rate (lfdr) framework with a prespecified threshold $\delta = 0.2$, following Efron [32], which provides an empirical Bayes approach to multiple testing. This framework has been shown to be optimal in the sense of controlling the false discovery rate in large-scale multiple testing problems [33]. For each taxon, we obtained posterior marginal distributions of $\beta_{A,j}$ and computed a z-score, defined as the posterior mean divided by the posterior standard deviation. This z-score serves as a standardized effect size that quantifies the strength of association, relative to posterior uncertainty. The lfdr values were then derived from these z-scores using the *locfdr* method within the *fdrtool* function in R. Taxa with lfdr values below a prespecified threshold $\delta$ were selected as significant features. Notably, lfdr does not require the test statistics (i.e., z-scores) to be independent, making it particularly suitable for correlated features in high-dimensional settings.

For ANCOM-BC and MaAsLin, feature selection was based on Benjamini-Hochberg (BH) adjusted q-values returned by the *ancombc2* and *maaslin3* functions, respectively, and taxa were selected using the significance level $\alpha = 0.1$. Although the two approaches are not identical–lfdr provides a Bayesian, posterior-probability-based criterion, whereas ANCOM-BC and MaAsLin use a frequentist multiple-testing correction–we compared them directly in the Results section to assess their relative effectiveness under the same data-generating scenarios. Notably, the lfdr threshold $\delta$ and the BH q-value threshold $\alpha$ arise from different statistical paradigms and are therefore not directly comparable. Following Efron [32], an lfdr threshold of $\delta = 0.2$ corresponds to q-values thresholds between 0.05 and 0.15, which motivated our choice of $\alpha = 0.1$ for ANCOM-BC and MaAsLin. We therefore interpret the results as empirical operating characteristics under method-specific selection rules, rather than as comparisons under identical nominal error-control criteria.

Based on the selected features from each method, we computed the true positive rate (TPR), defined as the number of true positives divided by the number of truly associated taxa, and false positive rate (FPR), defined as the number of false positives divided by the number of truly null taxa. These metrics were then averaged across simulation replicates.

For estimation accuracy for the main parameters of interests, $\beta_{A,j}$ and $\rho_w$, we evaluated the bias and mean square error (MSE) of the posterior means.

## FAMiLI Study

Fecal samples were self-collected, returned to NYU, and stored at -80°C. Microbial DNA was extracted using the PowerSoil kit, and the V4 region of the 16S rRNA gene was sequenced at Argonne National Laboratory. Raw sequencing reads were processed with QIIME2 [34] and DADA2 [35] to generate ASVs. Taxonomy was assigned using the Greengenes 2 reference database (October 2022 release), and a phylogenetic tree was constructed by inserting ASV sequences into the Greengenes reference phylogeny using the q2-frament-insertion plugin. Comprehensive cohort details are available in Kwak, Usyk [22].

### Lung Microbiome Study

We performed an analysis of lower airway microbiome data from a cohort of non-small cell lung cancer (NSCLC) patients. Enrolled participants were treatment-naïve NSCLC patients who underwent a clinically indicated diagnostic bronchoscopy at NYU Langone Health between 2018 and 2023. Institutional review board approval (NYU IRB s18-01845) and written informed consent were obtained from all patients prior to bronchoscopy. Lower airway samples were collected via bronchoalveolar lavage (BAL) from the tumor-involved lung segment. Taxonomic profiles were generated from metagenomic sequencing using KrakenUniq (v1.0.4) and Bracken (v2.8) [36, 37] run with default settings against the MicrobialDB reference database. Ambient air pollution exposure data were harmonized and obtained using the exact workflow deployed for the FAMiLI cohort [23].

## Results

### Simulation

Figure 3 illustrates the empirical TPR and FPR across Scenarios I-VI. Across all scenarios, the proposed model (SpaMixed) and MaAsLin achieved consistently high TPRs, exceeding 0.9 in all settings, whereas ANCOM-BC showed substantially lower sensitivity, with TPRs below 0.9 and decreasing as the number of taxa increased. This pattern indicates that ANCOM-BC behaved more conservatively and missed a larger proportion of truly associated taxa, particularly in higher-dimensional settings. In contrast, SpaMixed maintained high detection power across increasing values of J, demonstrating its ability to recover truly associated taxa under both moderate and strong spatial dependence.

The FPR results show a complementary pattern. ANCOM-BC maintained near-zero FPRs across all scenarios, reflecting its conservative selection behavior. SpaMixed also maintained low FPRs below the reference level of 0.05 in all scenarios, indicating that its high sensitivity was achieved without substantial inflation of false positives. MaAsLin also achieved high TPRs, but its FPR was elevated in Scenarios I and IV, where J=100, exceeding the 0.05 reference level. Thus, MaAsLin showed strong detection power but less stable false-positive control in some settings. Taken together, Figure 3 suggests that SpaMixed provides a favorable balance between sensitivity and false-positive control: it substantially improves TPR compared with ANCOM-BC while maintaining lower and more stable FPRs than MaAsLin in the scenarios where MaAsLin produced elevated false positives.

To further assess estimation accuracy, we compared bias and MSE of the main parameters of interest, $\beta_{A,j}$ and $\rho_w$. In Table 1, we present results for $J = 200$ under moderate spatial dependence, which is similar to the FAMiLI data. Results under other scenarios were provided in Additional file 1: Tables S3-S7. Across methods, SpaMixed showed the most favorable bias-variance trade-off. Specifically, SpaMixed exhibited a mean bias of -0.024 (truly significant taxa) and -0.025 (null taxa), with corresponding mean MSEs of 0.048 (truly significant taxa) and 0.040 (null taxa). In contrast, ANCOM-BC showed a mean bias of -0.015 (truly significant taxa) and -0.009 (null taxa), but higher mean MSEs (0.054 for truly significant taxa and 0.058 for null taxa) than SpaMixed. MaAsLin exhibited moderate bias and substantially larger MSE for truly significant taxa, indicating markedly less accurate effect size estimation. Concretely, the mean bias was -0.382 for truly significant taxa and 0.002 for null taxa, while the mean MSE was 3.203 and 0.002, respectively. Because MaAsLin fits feature-wise abundance models based on transformed nonzero relative abundances, it performs well for null taxa and can effectively rank true signals for variable selection. However, its effect size estimates for truly associated taxa remain substantially less accurate. Consequently, while MaAsLin distinguishes associated from null taxa reasonably well, it shows reduced accuracy in estimating the magnitude of the true exposure effects. In addition, the spatial dependence parameter $\rho_w$ also had small bias and MSE. These findings indicate that the spatial model improves feature selection performance, maintaining accurate coefficient estimation.

Under the additional negative-binomial scenario with moderate dependence and $J = 200$, we evaluated both variable selection performance and estimation accuracy for significant taxa. As shown in Additional file 1: Figure S3, all three methods achieved high TPRs, with MaAsLin showing the highest TPR (0.97), followed by SpaMixed (0.83) and ANCOM-BC (0.81). ANCOM-BC and MaAsLin showed slightly lower FPRs, whereas SpaMixed exhibited the highest FPR among the three methods in this setting; however, the FPR for SpaMixed remained below 0.05, indicating that false-positive selection was still well controlled. Thus, under this non-ZIP count-generating mechanism, SpaMixed maintained competitive variable-selection performance while preserving low false-positive rates. We further examined estimation accuracy for the truly significant taxa (Additional file 1: Tables S8). SpaMixed and ANCOM-BC both yielded small biases for the truly significant taxa, with average biases of -0.016 and 0.023, respectively, whereas MaAsLin showed a larger bias with average value of -0.382. A similar pattern emerged for MSE; SpaMixed and ANCOM-BC achieved low MSEs for the truly significant taxa (average MSEs of 0.119 and 0.091, respectively), while MaAsLin exhibited substantially higher MSEs (average MSE of 3.295). For null taxa, all methods produced estimates centered close to zero. MaAsLin had the smallest null-taxa MSE, while ANCOM-BC and SpaMixed also maintained low null-effect errors. These findings aligned with our primary results, confirming that while MaAsLin effectively ranks true signals for variable selection, its effect-size estimates for truly associated taxa are notably less accurate. For the spatial dependence parameter $\rho_w$, SpaMixed also produced small bias and MSE, further reflecting stable recovery of the underlying spatial dependence structure. Therefore, the negative-binomial robustness analysis suggests that SpaMixed remains competitive in feature selection, providing accurate estimation of exposure effects and stable spatial dependence recovery under a non-ZIP count distribution.

Under the sensitivity analysis for Scenario II (moderate dependence, J = 200) with M = 50 Monte Carlo simulations, we further evaluated the impact of prior specifications for the exposure effects and variance components. As shown in Additional file 1: Table S9, the baseline specification yielded the most balanced performance, achieving the highest TPR (0.94) with low FPR (0.02). Under alternative prior settings, modest reductions in TPR (0.88-0.9) were observed, while FPR remained stable at 0.03. These results suggest that variable selection performance is moderately sensitive to prior specification but remains stable in terms of false positive control. Consistent with these findings, the estimation accuracy results in Additional file 1: Table S10 show that the baseline setting yields substantially lower bias and MSE for the truly significant taxa compared with alternative prior specifications. This indicates that accurate effect size estimation is more sensitive to prior misspecification, reinforcing the choice of the baseline specification.

## FAMiLI Study

The proposed method was applied to the FAMiLI dataset. After excluding participants with missing covariate data, 359 individuals from 82 postal codes in NYC were included in the analysis. Participants ranged in age from 38 to 87 years (116 male and 243 female).

We fitted the proposed model, adjusting for age, gender, and body mass index. Continuous covariates were standardized (i.e., centered and scaled to unit variance) prior to model fitting. A Beta$(2,2)$ prior was assigned to the dependence parameter $\rho_w$. This prior allows moderate correlation among taxa, reflecting the biological expectation that related taxa are partially but not fully correlated in their effects. The remaining parameters used the same prior specifications as defined earlier.

The neighborhood matrix for the region-level random effect was constructed using Queen's case contiguity based on postal codes boundaries, with adjacency defined by shared borders or vertices. For the taxon-level random effect, we constructed binary neighborhood matrices based on two criteria: (i) functional similarity at the genus level, obtained by thresholding the functional similarity matrix at 0.1 (Additional file 1: Figure S4), and (ii) phylogenetic distance, using the 10th percentile of all pairwise distances as the threshold (Figure 4). For visualization, the matrices are displayed using their lower-triangular representation to avoid redundant arising from symmetric relationships.

Our analysis focused on the exposure effect, our primary variable of interest. Significant taxa were identified using the lfdr with a threshold of $\delta = 0.2$. Our proposed model identified 11 significant taxa; notably, this selection remained identical regardless of whether the neighborhood matrix was based on functional similarity or phylogenetic distance. To further assess concordance between methods, we compared the results with ANCOM-BC and MaAsLin. With threshold of $\alpha = 0.1$, ANCOM-BC and MaAsLin did not identify any taxa as significant after multiple-testing correction. To investigate this discrepancy, we examined the distribution of p-values for the exposure effect across ANCOM-BC and MaAsLin (Additional file 1: Figure S5). Although no taxa reached statistical significance, several taxa identified by our model were located in the lower tail of the p-value distributions. It suggests partial agreement in signal ranking despite differences in statistical significance thresholds.

We explored the spatial structure in the 11 taxa identified as significantly associated with $PM_{2.5}$. Following the same procedure described in the Motivating Example section, we performed a multivariate spatial autocorrelation test on the principal component scores obtained from PCA of CLR-transformed abundances restricted to these taxa. The test revealed a significant spatial structure (p-value = 0.001 based on 999 permutations), indicating that the reduced set of taxa retains coherent spatial structure. Figure 5(a) displays the spatial distributions of the average scores on the PC1 aggregated at the postal code level, highlighting clear geographic gradients. Figure 5(b) illustrates the association between the individual-level PC1 scores and $PM_{2.5}$ levels, demonstrating a consistent relationship between the dominant axis of microbial variation and $PM_{2.5}$ (Pearson correlation coefficient $R = 0.22$, $p$-value $< 0.05$). To enhance interpretability, we further examined the PC1 loadings to identify the taxa contributing most strongly to this axis of variation (Additional file 1: Figure S6). This analysis highlights the key microbial drivers underlying the observed relationship between PC1 and $PM_{2.5}$.

The 11 identified gut microbiota taxa exhibiting significant associations with $PM_{2.5}$ exposure are shown in Figure 6. Taxonomic labels in Figure 6 were defined according to the finest available level of taxonomic resolution to ensure clarity in the visualization. For ASVs resolved to the species level, binomial nomenclature (Genus species) was applied. For ASVs classified only to the genus level, the genus name was reported with the suffix 'sp.'. For ASVs that could not be resolved to the genus level, the lowest identified taxonomic rank was used, with ‘Uncl.’, indicating an unclassified lower rank, such as *CAG-138 Uncl.* Importantly, *Akkermansia* has been reported in studies of air pollution related microbiome alterations; this genus is crucial for gut barrier integrity and may be increased by $PM_{2.5}$ exposure [38]. Other genera, including several members of the families *Oscillospiraceae* (e.g., *CAG-138*, *CAG-177*, *Butyrivibrio*), showed significant associations to the $PM_{2.5}$ in our model, although direct evidence linking them to $PM_{2.5}$ is limited. While a subset of the identified genera (notably *Akkermansia*) have well-documented links to $PM_{2.5}$ exposure, many others may represent novel associations. These findings underscore the importance of gut microbial composition in understanding the biological impact of air pollution.

## Lung Microbiome Study

We further applied the proposed method to a lung microbiome dataset. The analysis focused on participants residing in NYC and included 84 individuals. Participants ranged in age from 50 to 88 years, with 20 male and 64 female included in the study. After filtering, taxa were retained if they had relative abundance greater than 0.1% in at least 10% of samples and raw counts greater than 30 in at least 10% of samples. This combined filtering criterion retained 46 taxa for the subsequent analysis. Because the lung microbiome dataset had a relatively small sample size (n=84), we used this strict filtering strategy to reduce instability from low-prevalence or low-count taxa and improve the stability of effect estimation. We then performed a multivariate spatial autocorrelation test on the principal components derived from the filtered taxa. The test provided evidence of spatial structure (p-value = 0.1 based on 999 permutations). Although this result did not reach the conventional 5% significance level, it suggests a possible spatial pattern in microbiome variation across locations. The spatial distributions of the average values of the first and second principal components are shown in Additional file 1: Figure S7.

We fitted the proposed model to the lung microbiome dataset while adjusting for age, gender, and smoking status. Prior to model fitting, all continuous covariates were standardized (i.e., centered and scaled to unit variance). The prior distributions and hyperparameter specifications were chosen to be consistent with those used in the FAMiLI Study analysis. To account for dependence among taxa, we incorporated a taxon-level random effect informed by phylogenetic relatedness. Specifically, binary neighborhood matrix was constructed based on phylogenetic distances between taxa. Two taxa were considered neighbors if their phylogenetic distance was less than or equal to the 10th percentile of all pairwise distances, thereby connecting taxa that were phylogenetically close.

We focused on the exposure effect, which was the primary interest in this analysis. Using the lfdr with a threshold of $\delta = 0.2$, the proposed model identified one taxon as significantly associated with the exposure. To further evaluate concordance with existing methods, we compared our results with those obtained from ANCOM-BC and MaAsLin. Using a threshold of $\alpha = 0.1$, neither ANCOM-BC nor MaAsLin identified any taxa as significant after multiple-testing correction. As in the first data analysis, we further examined the distributions of the p-values from ANCOM-BC and MaAsLin for the exposure effect (Additional file 1: Figure S8). The taxon identified by the proposed model had the smallest ANCOM-BC p-value among the evaluated taxa, although it did not reach statistically significance after multiple-testing correction. This finding suggests partial concordance between SpaMixed and existing methods in terms of signal prioritization.

The lung microbiota taxon identified as significantly associated with $PM_{2.5}$ exposure is shown in Figure 7. Because this taxon was classified only to the genus level, it is labeled as *Pseudomonas sp.* in the figure. Prior studies have shown that $PM_{2.5}$ can carry diverse bacterial communities, including taxa assigned to the genus *Pseudomonas* [39, 40]. In addition, experimental and epidemiologic studies further support the relevance of *Pseudomonas*-related signals in respiratory responses to particulate matter exposure [41]. Because our taxonomic annotation was resolved only to the genus level, these findings should be interpreted as genus-level biological support rather than species-level confirmation. Consequently, the identification of *Pseudomonas sp.* supports the biological plausibility of $PM_{2.5}$-associated shifts in lung microbial composition.

## Discussion

In this article, we developed a spatial mixed model to investigate the associations between microbiome and $PM_{2.5}$ exposure. To our knowledge, this is the first microbiome modeling framework that explicitly incorporates both spatial dependency at region-level and ecological dependency at taxon-level, providing a novel tool for disentangling complex ecological and environmental correlations. By accounting for both sources of random effects, our approach achieved high true positive rates while maintaining low false positive rates. Furthermore, the inclusion of taxon-level random effect allowed the model to capture biologically meaningful similarities among taxa while addressing correlation in microbiome data. Application to real data identified several genera that have previously been linked to air pollution and gut or respiratory health, as well as novel taxa whose associations may provide new insights into microbial contributions to $PM_{2.5}$-related health effects.

In the proposed framework, we used a parsimonious ZIP model instead of zero-inflated negative binomial models, which can accommodate both excess zeros and overdispersion. This choice was motivated by the sparsity and excess zeros in microbiome count data, while the hierarchical region-, sample-, and taxon-level random effects introduce heterogeneity in the Poisson mean and allow the marginal count distribution to exhibit extra-Poisson variability. Furthermore, adding a negative binomial dispersion component would introduce additional hyperparameters, potentially taxon-specific dispersion parameters, which can be difficult to identify jointly with zero-inflation probabilities and multiple structured random effects in sparse high-dimensional settings.

Therefore, the ZIP formulation was chosen to balance flexibility, identifiability, and computational feasibility.

In implementing our framework, it is important to account for several practical considerations. First, tools such as Moran's I or multivariate spatial autocorrelation tests provide an efficient diagnostic to assess the presence of spatial autocorrelation before model fitting. Second, the model includes dependence parameters that capture residual structure in the data. In particular, the spatial dependence parameter $\rho_u$ reflects similarity induced by spatially structured latent processes, while the taxon-level dependence $\rho_w$ captures shared structure among taxa. Together, these parameters provide a qualitative decomposition of spatial versus taxonomic contributions to residual variation. Careful attention is required for the taxon-level dependence parameter $\rho_w$. The choice of prior distribution for $\rho_w$ can be guided by biological knowledge. For instance, a Beta$(2{,}2)$ prior centers inference around moderate correlation, while a Beta$(5{,}1)$ prior reflects stronger expected dependence. This flexibility allows researchers to tailor the model to the ecological and biological context of the microbiome data under study. Third, we binarize either the phylogenetic distance matrix or functional similarity matrix to construct a sparse adjacency matrix for computational stability and model parsimony. This binarization defines a local dependence structure in which taxa within a specified biological distance threshold are treated as neighbors, while more distant taxa are assumed conditionally independent. The resulting sparse neighborhood matrix is compatible with the structured taxon-level random effect and improves scalability in high-dimensional microbiome data. Although this work uses binary adjacency matrices, the framework can also accommodate continuous distance- or similarity-based weights, such as distance-decay functions, when such weighting is biologically justified. However, continuous weights require additional choices regarding scaling or bandwidth and may increase computational complexity. Fourth, the exposure was generated using a CAR model to induce spatial dependence in simulation studies. However, the proposed method does not rely on any specific distributional assumption on the exposure process. In particular, it does not require the exposure to follow a CAR model and is directly applicable to exposures generated from alternative spatial processes such as simultaneous autoregressive models. Additionally, in applications, the region-level random effect was defined using postal codes as the spatial units. However, the proposed framework can accommodate other forms of spatial information, such as geographic coordinates, as long as an appropriate spatial neighborhood matrix can be constructed.

One strength of our proposed framework is its adaptability to a variety of biologically motivated similarity measures of taxon-level dependence. While we demonstrated the use of phylogenetic and functional similarity-based distances, other metrics such as co-occurrence correlation networks [10, 42], ecological niche similarity [11], or metabolic interaction-based distances [43, 44] can be readily incorporated. This extensibility highlights the broad applicability of our approach to diverse microbiome research questions, particularly in environmental health contexts where multiple ecological relationships may shape microbial responses to exposures [45, 46]. Because different similarity measures can lead to different dependence structures, we view the choice of taxon-level dependence measure as a sensitivity or user-specified modeling decision rather than a single universally optimal definition. Indeed, benchmark studies have shown that microbial association-network methods can vary substantially in sensitivity and precision, especially under compositionality, uneven sequencing depth, rare taxa, and zero inflation [47]. Accordingly, our R package is designed to allow users to construct or supply alternative taxon similarity matrix, enabling the dependence structure to be tailored to the data source and biological question. In terms of computational efficiency, the INLA implementation is notably fast; for example, the proposed model for the FAMiLI study analysis took approximately 1.5 minutes to run, using a single core of an Apple M3 Pro (11-core CPU, 18 GB RAM). This runtime indicates that our framework is computationally practical for implementation. The R package is publicly available on Github (https://github.com/srkim3487/SpaMixed).

Beyond environmental exposure-microbiome studies, the proposed framework is broadly relevant to other high-dimensional count data measured on spatially indexed subjects. A substantial body of existing work has developed to address spatial autocorrelation, including spatial generalized linear mixed models and Gaussian process models, which have been widely applied in ecology, geostatistics, and environmental statistics [48]. These approaches are effective for modeling spatial dependence, but are typically not designed for high-dimensional molecular outcomes, where the number of features can be large and complex dependence structures exist both across space and features. In this context, our framework complements existing spatial modeling strategies by enabling computationally efficient inference for spatially indexed high-dimensional count data while incorporating both spatial structure and feature level dependence. This provides a bridge between classical spatial statistical methodology and modern high-throughput count-based biological data analysis.

Despite these strengths, several limitations should be noted. First, we focused on a single environmental exposure, $PM_{2.5}$. However, $PM_{2.5}$ can originate from diverse sources such as traffic emissions, soil dust and oil combustion, each of which may exert distinct biological effects. In addition, $PM_{2.5}$ may be correlated with unmeasured spatially varying factors, such as socioeconomic conditions and demographic composition. Including those factors raise the possibility of spatial confounding, whereby spatially structured exposure effects may be partially indistinguishable from latent spatial processes, potentially leading to biased effect estimates [49, 50]. Extending the framework to simultaneously incorporate multiple environmental exposures and additional spatial factors, while appropriately addressing spatial confounding, would provide a more nuanced understanding of environmental influences. Second, although our framework accommodates various taxon-taxon similarity structures, the selection of the most biologically meaningful metric remains an open question. In particular, the results suggest that taxon-level dependence is more weakly identified than regional spatial structure and is sensitive to prior specification. This highlights the challenges in characterizing fine-scale taxonomic features, whereas feature selection results remain consistent across a range of prior choices, indicating that the main inferential conclusions are robust. Future work incorporating additional validation using experimental or longitudinal data may help better characterize appropriate taxon-level dependency structures. Third, in the current formulation of our model, each subject is associated with a single region index. Therefore, the spatial structure is defined under a hard assignment setting, where each observation is linked to one region-specific random effect. An interesting extension of the proposed framework is to incorporate mobility-based spatial structures, where subjects may have partial membership across multiple regions, for example as a weighted combination of region-level effects. This would allow modeling more complex spatial exposure patterns, such as time-varying residence.

## Conclusions

Linking environmental exposures to the microbiome is central to understanding how they affect human health. However, existing models often overlook two key features of microbiome data: spatial variation across regions and correlations among microbial taxa. Our spatial mixed framework explicitly models both sources of dependence, thereby improving the accuracy and interpretability of feature selection and association inference. To our knowledge, this is the first microbiome modeling framework to incorporate both region-level spatial effects and taxon-level ecological dependence.

## List of abbreviations

$PM_{2.5}$: particulate matter; ZIP: zero-inflated Poisson; ANCOM-BC: analysis of compositions of microbiomes with bias correction; CAR: conditional autoregressive; INLA: Integrated nested Laplace approximation; MCMC: Markov Chain Monte Carlo; FAMiLI: Food and Microbiome Longitudinal Investigation; NYC: New York City; ASV: Amplicon Sequence Variants; OTU: operational taxonomic unit; CLR: centered log-ratio; PCA: Principal component analysis; PC1: first principal component; PC2: second principal component; GCN: genomic content network; MAP: maximum a posterior; PC: penalized complexity; ANCOM: analysis of compositions of microbiomes; lfdr: local false discovery rate; BH: Benjamini-Hochberg; TPR: true positive rate; FPR: false positive rate; MSE: mean square error; NYU: New York University; QIIME: quantitative insights into microbial ecology; DADA: Divisive Amplicon Denoising Algorithm; NSCLC: non-small cell lung cancer; BAL: bronchoalveolar lavage.

## Declarations

### Ethics approval and consent to participate

Not applicable.

### Consent for publication

Not applicable.

### Availability of data and material

The FAMiLI dataset have been deposited in the Sequence Read Archive (PRJNA559143), along with demographic metadata Kwak, Usyk [22]. The Lung Microbiome Study dataset have been deposited in Sequence Read Archives PRJNA1284602. R package used for the analyses presented in the manuscript is available at https://github.com/srkim3487/SpaMixed.

**Competing interests**

The authors declare that they have no competing interests.

**Funding**

The work was supported by NIH grants R01LM014085(SK[1], HL, LNS), U24ES036002(JA), R01HL178710(HL, SK[2], JA), R01CA159036 (SK[2], JA), U01CA250186(SK[2], JA), R33 GM147800 (HL, LNS, NIGMS/NIH); U01 AG088351 (LNS, NIA/NIH); R37 CA244775 (HL, LNS, NCI/NIH); U2C CA271890 (LNS, NCI/NIH)

**Authors' contributions**

S.K[1] designed research, performed all analyses, and wrote the paper; C.W contributed to statistical modeling, real data analysis and revised the paper; F.D, A.B, L.N.S, S.K[2,3] and J.A provided real data and biological insight and revised the paper; H.L conceived and designed the study, supervised research, and wrote the paper.

**Acknowledgments**

Not applicable.

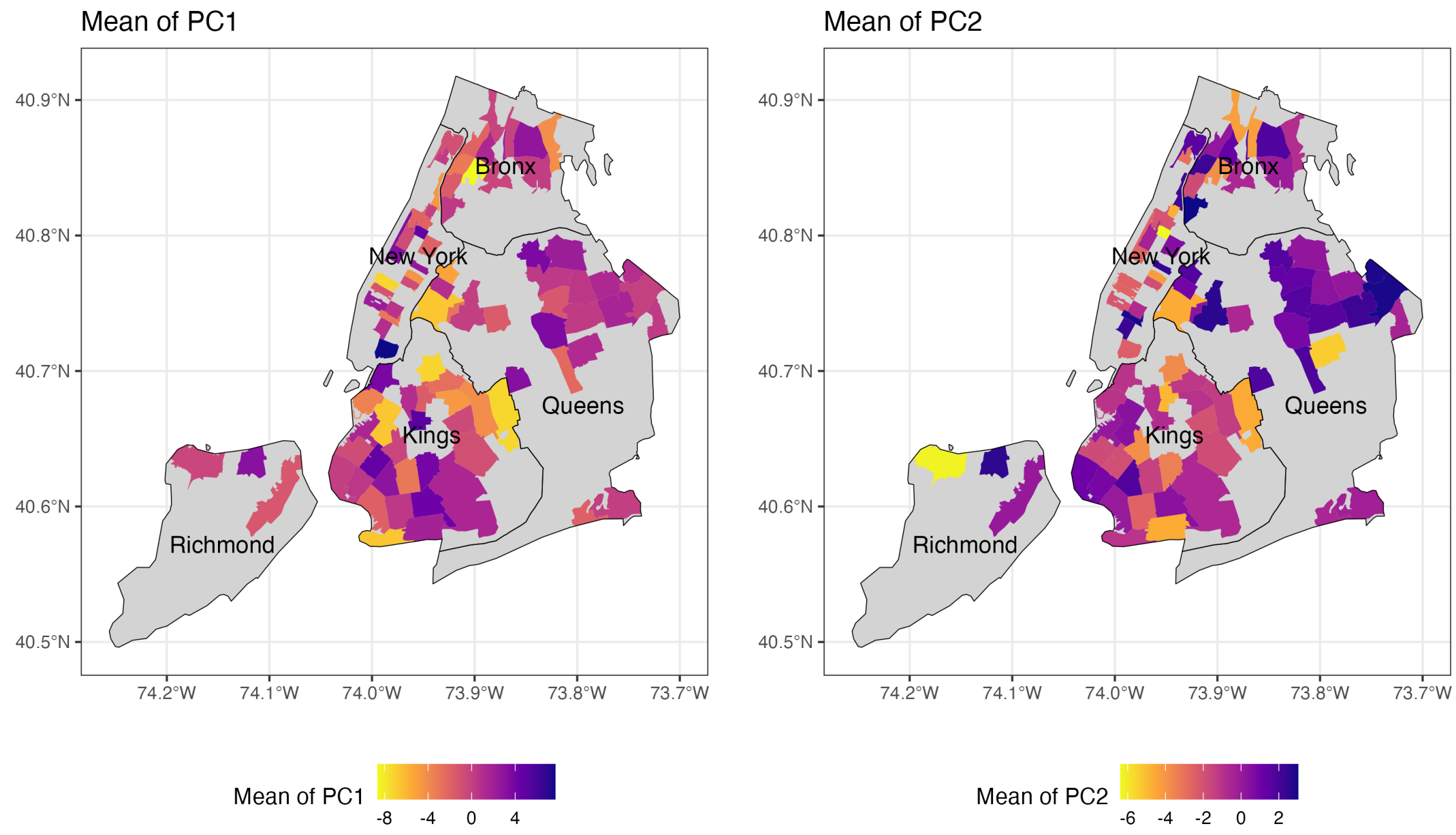

**Figure 1. Spatial distributions of average microbiome principal component scores across New York City postal codes (FAMiLi dataset).** The left and the right panels show postal code-level mean scores on the first and second principal components, respectively, derived from PCA of CLR-transformed taxonomic abundances.

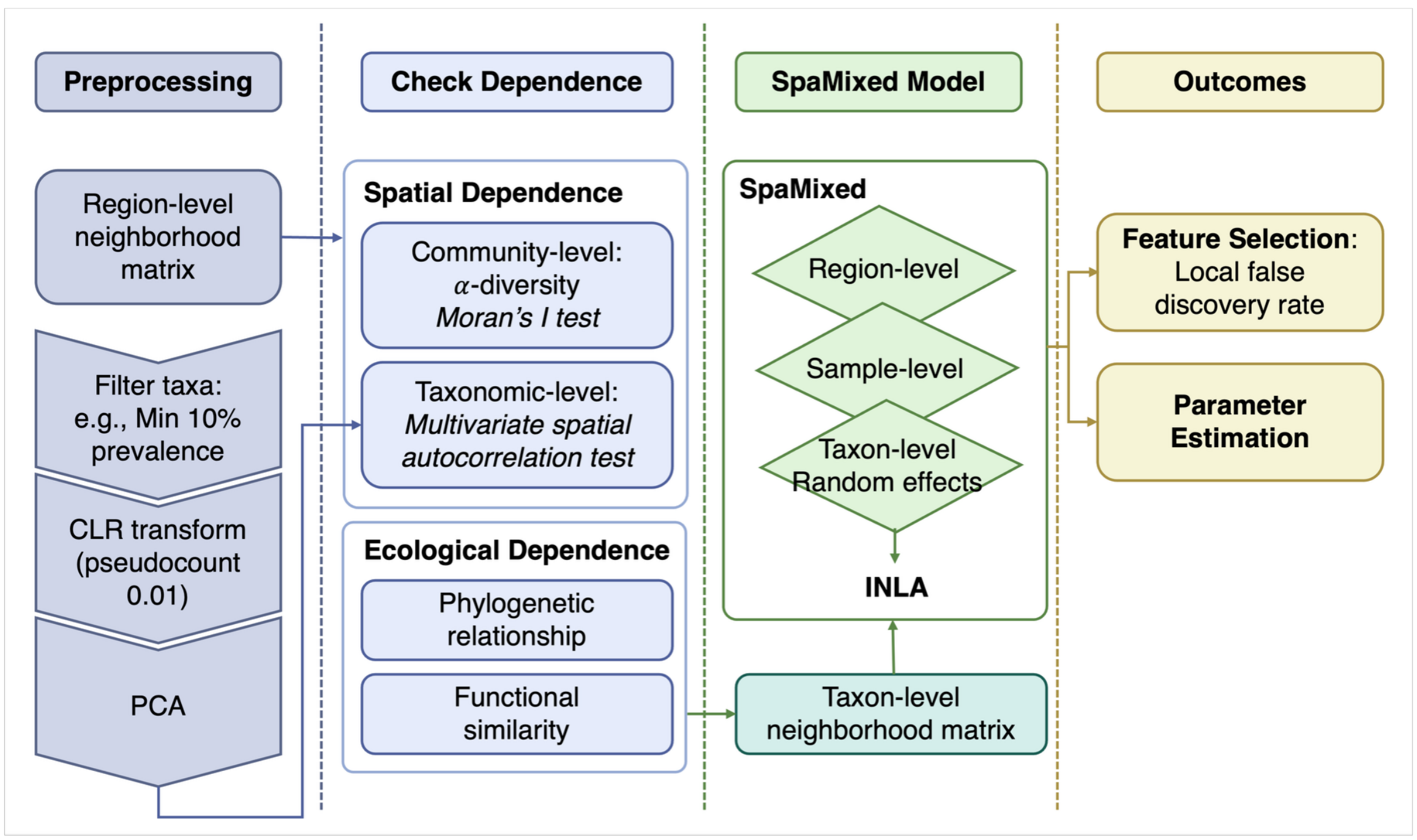

**Figure 2. Workflow of the proposed framework.** The pipeline begins with data preprocessing, followed by assessment of spatial and ecological dependencies. If such dependencies are present, the proposed method, SpaMixed, is applied. Parameter estimation is conducted under a Bayesian method, INLA, with appropriate prior specifications and shrinkage structures. Feature selection is then performed using local false discovery rate control.

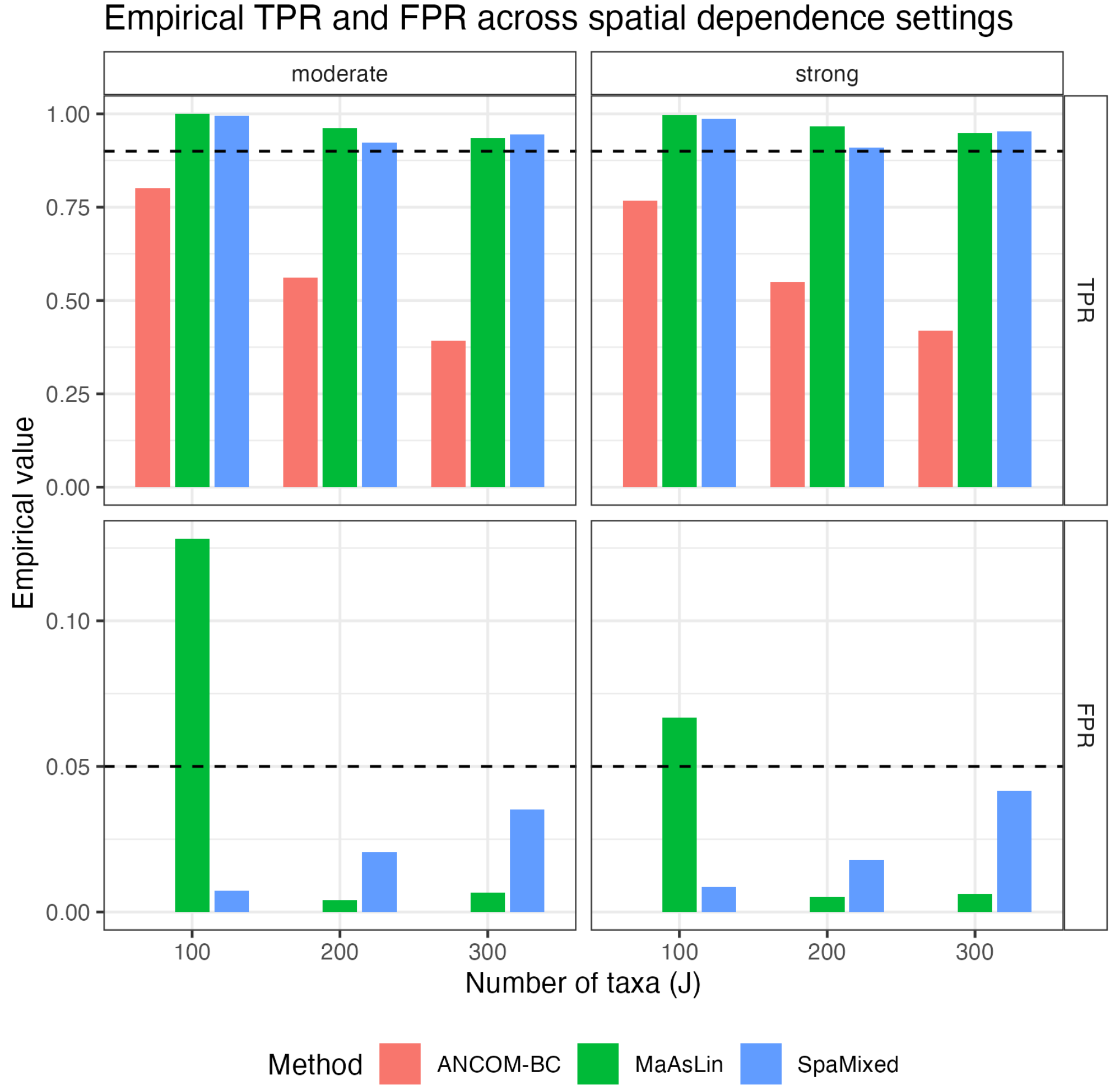

**Figure 3. Empirical true positive rate and false positive rate across simulation scenarios.** Bars show the average empirical true positive rate (TPR) and false positive rate (FPR) across simulation replicates for SpaMixed, ANCOM-BC, and MaAsLin. Columns correspond to moderate and strong spatial dependence settings, and rows correspond to TPR and FPR. The x-axis indicates the number of taxa, J=100,200, or 300. SpaMixed selected features using the lfdr threshold $\delta = 0.2$, whereas ANCOM-BC and MaAsLin used the BH-adjusted p-value threshold $\alpha = 0.1$. For MaAsLin, abundance-model results were used to align with the simulated exposure effects in the count-abundance component. Dashed horizontal reference lines indicate TPR = 0.90 and FPR = 0.05.

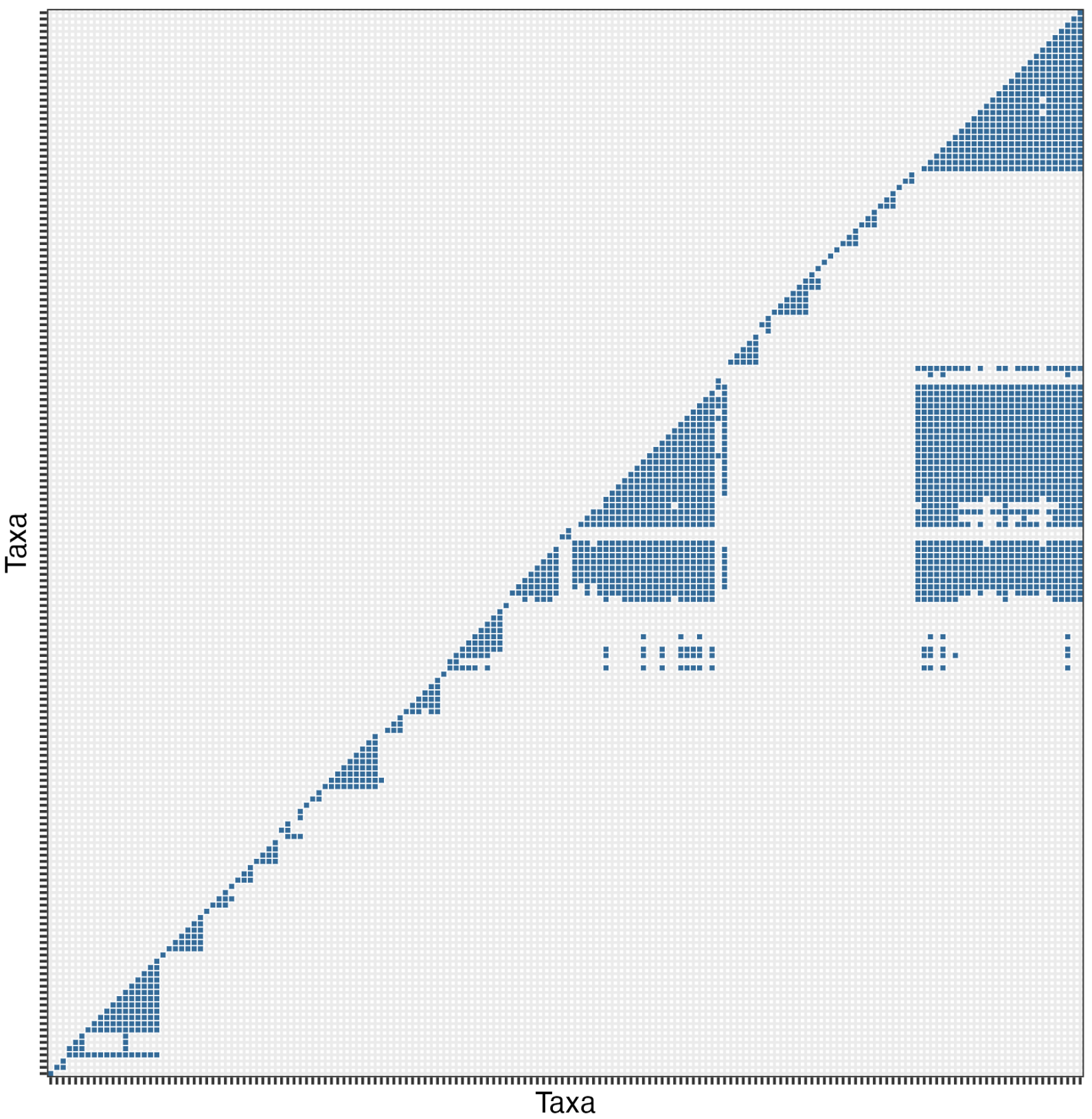

**Figure 4. Binary neighborhood matrix based on phylogenetic distance (FAMiLi dataset).** The matrix represents phylogenetic connectivity between taxa, where a value of 1 (blue) indicates that a pairwise cophenetic distance is less than or equal to the 10th percentile of all pairwise distances, and 0 (white) indicates no connection.

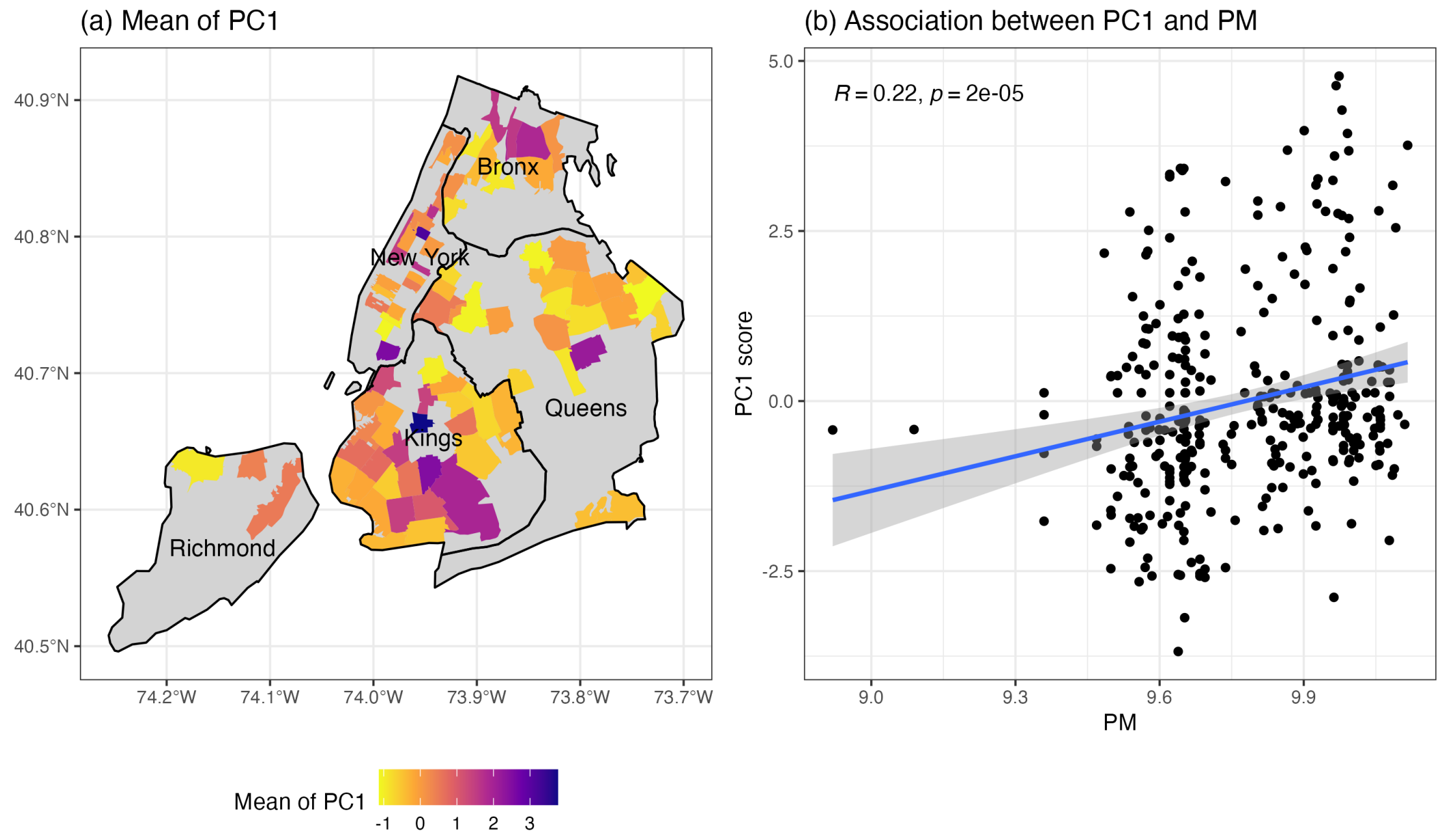

**Figure 5. Geographic distribution and environmental relationship of identified taxa in the FAMiLi dataset.** (a) Spatial distribution of postal code level mean PC1 scores, derived from PCA of CLR-transformed abundances restricted to the 11 identified taxa identified by the SpaMixed model. (b) Association between PC1 scores and $PM_{2.5}$ levels. Each point represents an individual observation. The solid line indicates the fitted linear regression, with the shaded band representing the 95% confidence interval. Pearson correlation coefficient $(R)$ and the associated p-value $(p)$ are provided to quantify the strength of the relationship.

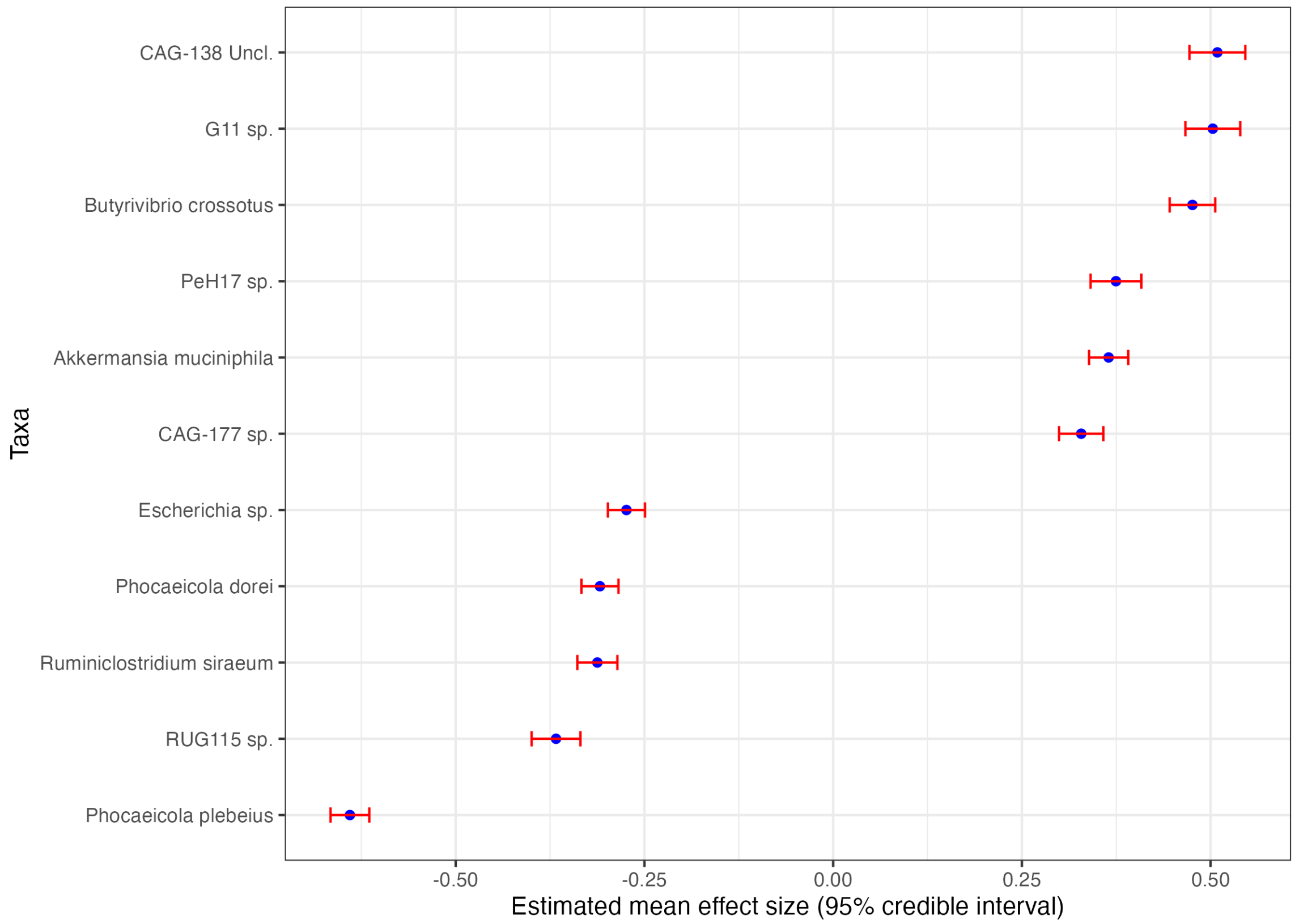

**Figure 6. Taxa-specific effect estimates for FAMiLi dataset derived from the SpaMixed model.** Forest plot displays posterior mean estimates and corresponding 95% credible intervals for the effect of $PM_{2.5}$ on significantly associated gut microbial taxa. Blue points represent posterior mean effect sizes, while red horizontal bars denote the 95% credible intervals (2.5%–97.5% quantiles). Taxa are ordered along the y-axis by the magnitude of the posterior mean for visualization clarity. Taxonomic labels were defined according to the finest available level of taxonomic resolution to ensure clarity in the visualization. For ASVs resolved to the species level, binomial nomenclature (Genus species) was applied. For ASVs classified only to the genus level, the genus name was reported with the suffix 'sp.'. For ASVs that could not be resolved to the genus level, the lowest identified taxonomic rank was used, with ‘Uncl.’, indicating an unclassified lower rank, such as CAG-138 Uncl.

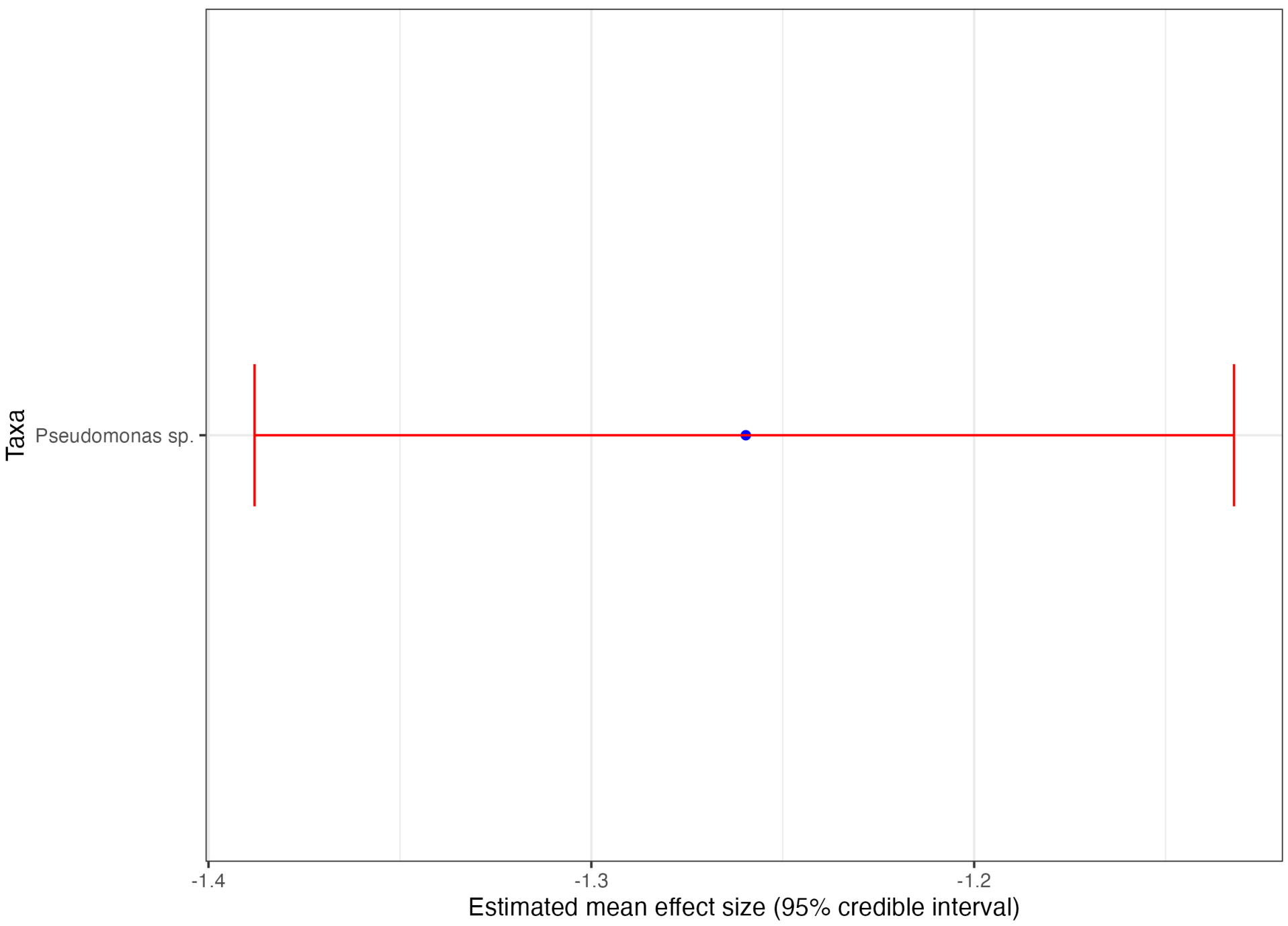

**Figure 7. Taxon-specific effect estimate for lung microbiome dataset derived from the SpaMixed model.** Plot displays posterior mean estimates and corresponding 95% credible intervals for the effect of $PM_{2.5}$ on significantly associated lung microbial taxa. Blue points represent posterior mean effect sizes, while red horizontal bars denote the 95% credible intervals (2.5%–97.5% quantiles).

**Table 1. Bias and MSE of the posterior means for significant taxa $\beta_{A,j}$ and $\boldsymbol{\rho_w}$ under Scenario II (moderate, J=200).**

| | Bias | | | MSE | | |
|---|---|---|---|---|---|---|
| | ANCOM-BC | MaAsLin | SpaMixed | ANCOM-BC | MaAsLin | SpaMixed |
| $\beta_{A,123}$ | -0.07 | -2.05 | -0.05 | 0.03 | 4.20 | 0.02 |
| $\beta_{A,129}$ | -0.03 | -1.21 | -0.05 | 0.02 | 1.48 | 0.02 |
| $\beta_{A,133}$ | 0.06 | 1.33 | -0.05 | 0.02 | 1.77 | 0.05 |
| $\beta_{A,134}$ | -0.08 | -2.16 | -0.06 | 0.02 | 4.68 | 0.02 |
| $\beta_{A,135}$ | 0.06 | 2.73 | 0.04 | 0.05 | 7.47 | 0.04 |
| $\beta_{A,136}$ | -0.04 | -1.61 | -0.05 | 0.04 | 2.59 | 0.02 |
| $\beta_{A,137}$ | 0.04 | 1.79 | 0.01 | 0.02 | 3.21 | 0.02 |
| $\beta_{A,138}$ | -0.06 | -1.95 | -0.05 | 0.02 | 3.83 | 0.02 |
| $\beta_{A,139}$ | -0.09 | -1.42 | -0.20 | 0.13 | 2.02 | 0.12 |
| $\beta_{A,140}$ | -0.04 | -1.22 | -0.06 | 0.08 | 1.50 | 0.04 |
| $\beta_{A,141}$ | 0.08 | 1.57 | 0.25 | 0.16 | 2.49 | 0.17 |
| $\rho_w$ | - | | -0.1 | - | | 0.01 |

## Additional material

**Additional file 1: Table S1. Distribution of the number of subjects across locations**.

**Additional file 1: Table S2. Parameter settings for $\beta_A$ in the simulation studies.**

**Additional file 1: Table S3. Bias and MSE of the posterior means for significant taxa $\beta_{A,j}$ and $\rho_w$ under Scenario I (moderate, J=100).**

**Additional file 1: Table S4. Bias and MSE of the posterior means for significant taxa $\beta_{A,j}$ and $\rho_w$ under Scenario III (moderate, J=300).**

**Additional file 1: Table S5. Bias and MSE of the posterior means for significant taxa $\beta_{A,j}$ and $\rho_w$ under Scenario IV (strong, J=100).**

**Additional file 1: Table S6. Bias and MSE of the posterior means for significant taxa $\beta_{A,j}$ and $\rho_w$ under Scenario V (strong, J=200).**

**Additional file 1: Table S7. Bias and MSE of the posterior means for significant taxa $\beta_{A,j}$ and $\rho_w$ under Scenario VI (strong, J=300).**

**Additional file 1: Table S8. Bias and MSE of the posterior means for significant taxa $\beta_{A,j}$ and $\rho_w$ under Scenario II (moderate, J=200), based on microbiome counts generated from negative-binomial distribution.**

**Additional file 1: Table S9. Comparison of variable selection performance under sensitivity analysis for Scenario II (moderate, J=200), based on $M = 50$ Monte Carlo simulations.**

**Additional file 1: Table S10. Bias and MSE of the posterior means for significant taxa $\boldsymbol{\beta}_{A,j}$ under sensitivity analysis for Scenario II (moderate, J=200), based on $M = 50$ Monte Carlo simulations.**

**Additional file 1: Figure S1. Spatial distribution of microbial $\alpha$-diversity across the FAMiLi dataset.** Maps represent the geographic distribution of average values for four distinct $\alpha$-diversity metrics: Observed ASVs, Shannon index, Simpson's index, and Inverse Simpson index. Data points are aggregated by postal code.

**Additional file 1: Figure S2. Phylogenetic and functional relationships among 212 taxa in the FAMiLi dataset.** Heatmaps represent (a) the cophenetic distance matrix derived from the phylogenetic tree and (b) the functional similarity matrix. Each cell represents the pairwise value between two taxa, with colors indicating the magnitude of phylogenetic distance and functional similarity, respectively.

**Additional file 1: Figure S3. Empirical true positive rate and false positive rate under Scenario II (moderate dependence, J=200), based on microbiome counts generated from negative-binomial distribution.** Bars show the average empirical true positive rate (TPR) and false positive rate (FPR) across simulation replicates for SpaMixed, ANCOM-BC, and MaAsLin. Rows correspond to TPR and FPR. SpaMixed selected features using the lfdr threshold $\delta = 0.2$, whereas ANCOM-BC and MaAsLin used the BH-adjusted p-value threshold $\alpha = 0.1$. For MaAsLin, abundance-model results were used to align with the simulated exposure effects in the count-abundance component.

**Additional file 1: Figure S4. Binary neighborhood matrix of functional similarities for the FAMiLi dataset.** The matrix illustrates functional connectivity at the genus level, where a cell is defined as 1 if $1 - d_{ik} > 0.1$, and 0 otherwise ($d_{ik}$ denotes functional distance between taxon $i$ and taxon $k$). Blue cells indicate a value of 1, representing a functional connection between taxa, while white cells indicate no connection (0).

**Additional file 1: Figure S5. Rank distributions of p-values for ANCOM-BC and MaAsLin for the $PM_{2.5}$ effect in the FAMiLi dataset.** The plot displays the p-value distribution for the (left) ANCOM-BC and (right) MaAsLin analyses regarding the $PM_{2.5}$ effect. Grey points represent the full distribution of tested taxa, while red points highlight the specific taxa identified as significant by the SpaMixed model.

**Additional file 1: Figure S6. Loadings for the first principal component (PC1) in the FAMiLi dataset.** Bar plot displaying the loadings of the 11 taxa contributing to PC1, derived from PCA of CLR-transformed taxonomic abundances. Positive and negative loadings indicate taxa that are positively or negatively associated with the dominant community gradient captured by PC1.

**Additional file 1: Figure S7. Spatial distributions of average microbiome principal component scores (lung microbiome dataset).** The left and the right panels show scores on the first and second principal components, respectively, derived from PCA of CLR-transformed taxonomic abundances.

**Additional file 1: Figure S8. Comparative rank distribution of p-values for ANCOM-BC and MaAsLin across lung microbiome dataset.** The plots illustrate the p-value distributions for the (left) ANCOM-BC and (right) MaAsLin analyses regarding the $PM_{2.5}$ effect. Grey points represent the full distribution of tested taxa, while red points highlight the specific taxa identified as significant by the SpaMixed model.